\documentstyle[prd,aps,epsf,epsfig]{revtex} 
\begin{document}
\draft
\twocolumn[\hsize\textwidth\columnwidth\hsize\csname@twocolumnfalse\endcsname
%
\title{Self-organized criticality in
stick-slip models with periodic boundaries}
%
\author
{Kwan-tai Leung${}^1$,
J\o rgen Vitting Andersen${}^2$ and
Didier Sornette${}^{3,4}$}
\address{
${}^1$ Institute of Physics, Academia Sinica,
Taipei, Taiwan 11529, R.O.C.\\
${}^2$Department of Mathematics, Imperial College, Huxley Building,
180 Queen's Gate, London SW7 2BZ, U.K. \\
${}^3$ Laboratoire de Physique de la
Mati\`{e}re Condens\'{e}e, CNRS,
Universit\'{e}\ de Nice-Sophia Antipolis, \\
Parc Valrose, 06108 Nice Cedex 2, France\\
$^4$ Department of Earth and Space Sciences and Institute of Geophysics
and Planetary Physics\\
University of California, Los Angeles, California 90095-1567, U.S.A.
}

\maketitle
\centerline{\small (Last revised \today)}

\begin{abstract}

A spring-block model governed by threshold dynamics and
driven by temporally increasing spring constants is investigated.
Due to its novel multiplicative driving, criticality
occurs even with periodic boundary conditions via a
mechanism distinct from that of previous models.
This mechanism is dictated by a coarsening process.
The results show a high degree of universality.
The observed behavior should be relevant to a class of
systems approaching equilibrium via a punctuated threshold dynamics.

\end{abstract}

\vspace{2pc}
]


Out-of-equilibrium driven systems with threshold dynamics exhibit a rich
phenomenology, from synchronized behavior \cite{Strogatz,Hopfield} to
self-organized criticality (SOC) \cite{btw,ofc,Others}. SOC refers to the
spontaneous organization towards a kind of dynamical critical steady
state.
Threshold out-of-equilibrium dynamics encompasses many systems, such as
neural
networks, solid friction, rupture with healing, earthquakes and avalanches.
It is now understood qualitatively that there is a class of models
exhibiting SOC as
a result of their tendency to synchronize \cite{Threshold}. This tendency
is however
frustrated by constraints such as open boundary conditions \cite{Threshold}
and
quenched disorder \cite{fault} which leads to a dynamical regime at the
edge of
synchronization, the SOC state. Another class, so-called extremal models,
are
understood to exhibit SOC due to the competition between local strengthening
and
weakening due to interactions \cite{Paczuski}.
In a third class of models, SOC
results from  the tuning of the order parameter of a system exhibiting a
genuine
critical point to a vanishingly  small, but positive value, thus ensuring
that the
corresponding control parameter lies exactly at its critical value for the
underlying depinning transition \cite{Sor}. The issue is furthermore
complicated by
the fact that a notable fraction of numerical and experimental works
\cite{Cannelli,Clauss,Barkhausen} claiming the observation of SOC from the
measurements of powerlaw distributions rely on the slow sweeping of
a control parameter towards a critical point \cite{Sweeping,Sethna}.

The purpose of this letter is to present a variation of spring-block models
using a novel form of driving by {\it multiplicative}
loading. The surprising result is that, when the dynamics is conservative
and for
{\it periodic boundary conditions\/} (PBC), the system self-organizes into a
critical state with long-range order and power-law distributions.
This is in contrast to all previous stick-slip models
\cite{ofc,Threshold,Vilotte}
that need either open boundary conditions and/or
dissipation to exhibit SOC (in order for synchronization
{\it not\/} to occur).  None of the
four above mechanisms seem at work here. We conjecture
that
this new class appears as a result of the multiplicative nature of the
driving,
known in simpler models to lead to stationary dynamical states with power
law
distributions \cite{Solomon}. Our model exhibits the dual property of
approaching (without ever reaching) an equilibrium state in terms of block
positions while at the same time being characterized by a steady state in
the force variables.
To our knowledge our model is the first of its kind to
show SOC in approach to an equilibrium state, and as such
suggests a new class of experimental systems which could
exhibit SOC states.

\underline{\sl Model:}
We consider a two-dimensional spring-block model consisting
of an array of blocks
interconnected among nearest neighbors by coil springs.
The springs have the same
spring constant $K$ and relaxed spring length $l$.
Initially, the array is stretched to a lattice spacing $a>l$ and
placed on a frictional substrate which is characterized
by a static threshold $F_s$ for slipping.
Disorder is introduced in the form of
random displacements $(x,y)$ of the blocks about
the coordinates $(ia, ja)$ on a square lattice,
where $-A \leq x,y \leq A$ and $i,j=1,\dots,L$.
The force between two neighboring blocks at $\vec{r}$
and $\vec{r'}$ is given by the Hooke's law $( |\vec{r}-\vec{r'}|-l)K$.
Since we are interested in dynamics primarily governed by tensile stresses,
this nonlinear dependence on the coordinates
leads to unnecessary complications in the algorithm.
To simplify and compare with similar models, we
expand the expression to first order in $(x,y)$ to obtain
the force components on a block in the bulk at $(i,j)$ \cite{gofc}
\begin{eqnarray}
F^x_{i,j} &=&
(x_{i+1,j} + x_{i-1,j}-2 x_{i,j})K
\nonumber\\
&&\quad
+(x_{i,j+1} + x_{i,j-1}-2 x_{i,j})sK
\nonumber\\
F^y_{i,j} &=&
(y_{i+1,j} + y_{i-1,j}-2 y_{i,j})sK
\nonumber\\
&&\quad
+(y_{i,j+1} + y_{i,j-1}-2 y_{i,j})K,
\label{force}
\end{eqnarray}
where $s=1-l/a>0$ is the initial strain. It is important to stress that
the terms proportional to $s$
leads to anisotropic couplings
to nearest neighbors in the transverse direction.
The coupling and the SOC state disappears for $s = 0$ or in 1D-chains.

Since the forces are linear in displacements, it is possible
to invert (\ref{force}) and formulate the model solely in terms of force
variables, as in \cite{ofc}.
Starting with a stable configuration with
net force $F\equiv \sqrt{{F^x}^2+{F^y}^2}<F_s$ for all the blocks,
we drive the system by gradually increasing $K$\cite{footnote1}
until one of the blocks becomes unstable,
i.e., $K$ is increased to $K F_s/F_{\rm max}$ during loading,
where $F_{\rm max}$ denotes the spatial maximum of $F$ in the stable
configuration.
As in \cite{ofc}, the block is assumed to slip to
its equilibrium position defined by $F=0$, ignoring overshot:
\begin{eqnarray}
&&F^x_{i,j} \to 0, \quad
F^y_{i,j}\to 0; \nonumber\\
&&F^x_{i\pm 1,j} \to F^x_{i\pm 1,j}+ \alpha F^x_{i,j}, \quad
F^y_{i\pm 1,j} \to F^y_{i\pm 1,j}+ s\alpha F^y_{i,j};\nonumber\\
&&F^x_{i,j\pm 1} \to F^x_{i,j\pm 1}+ s\alpha F^x_{i,j}, \quad
F^y_{i,j\pm 1} \to F^y_{i,j\pm 1}+ \alpha F^y_{i,j},
\label{slip}
\end{eqnarray}
where $\alpha=1/(2+2s)$.
This locally conserves the force components, thus the conservative nature
of the model.
The resulting modification of the stress environment may trigger
further slips in neighboring blocks, and hence an avalanche,
until $F<F_s$ is restored for all blocks.
Then $K$ is increased again and the slip process continues.

If $F_s$ was zero, the only stable (minimal energy) configuration would be
that the
blocks were exactly at the nodes of a perfect square
lattice of mesh size $a$.
The non-vanishing friction thus creates a large ensemble of coexisting
metastable states which is responsible for the
nontrivial dynamics. For $F_s \neq 0$
and $s \neq 0$, the toppling rules (\ref{slip}) do not put the blocks in
their minimum energy configuration due to the couplings to
their four neighbors.
This ensures that a block will go on becoming unstable
ad infinitum as long as $K$ is increased indefinitely.

The {\em multiplicative\/} loading
is motivated by the stiffening of an overlayer
caused by desiccation\cite{expt},
originally used to
study cracks\cite{abj}. It differs from the usual {\em additive\/} loading
in sandpiles \cite{btw,ofc,Others,Threshold} and
stick-slip models\cite{Vilotte,gofc}
where physically the driving force arises from
the dropping of grains onto a pile or
from the steady relative motion of two frictional surfaces.
Those systems are known to exhibit
self-organized criticality in non-equilibrium steady states.
In contrast, our system approaches an {\em equilibrium}
instead of a
genuine steady state.

Without loss of generality, we hereafter
set
$a=1=F_s$.
Of the remaining dimensionless parameters $\{ s,A,K \}$,
$s$ determines the equilibrium length scale
and the dynamics through (\ref{slip}),
$A$ characterizes the initial disorders
but is irrelevant for the equilibrium state,
and $K$ defines the ``time'' $t\equiv K$.

\underline{\sl Results:}
We have simulated the model for
$0\leq s\leq 1$, $A=0.02, 0.18, 0.4$,
and system sizes $20\leq L \leq 300$.
We are mainly interested in the possibility of SOC with PBC as this most
differentiates us from previous works.
Unless stated otherwise, PBC is assumed hereafter.
To investigate the effects of spatial inhomogeneities,
we also use
free boundary conditions (FBC) with no block beyond the edges and corners,
and cylindrical boundary conditions (CBC)
with one pair of parallel edges periodic and the other pair free.
All these boundary conditions respect the conservation law of the force
component.

The evolution of the system is monitored by the variation of
the stress field $\sigma(\vec{r},t)$, approximated by
the averaged tension of the four springs attached to a block.
Using its Fourier transform $\tilde \sigma$,
we compute the structure factor
${\cal S}(\vec{k},t) = \langle |\tilde \sigma(\vec{k},t)|^2 \rangle/L^2
-L^2 \delta_{\vec{k},0}
\langle\overline{\tilde\sigma(\vec{k},t)}\rangle^2$
and the circular average
${\cal S}_{\rm cir}(k,t)= \sum_{\vec{k}\in\{|\vec{k}|=k\}}
{\cal S}(\vec{k},t)$,
where the overline and the angular brackets mean a spatial and
an ensemble average, respectively.
Then $R(t)\equiv \left[ \sum_{k} k^2 {\cal S}_{\rm cir}(k,t)/
\sum_{k} {\cal S}_{\rm cir}(k,t)\right]^{-1/2}$
gives a simple measure of the characteristic length for the stress field.

The system evolves under a conservative dynamics from large disorders to
small disorders when the blocks
converge onto a perfect lattice.
This is analogous to spinodal decomposition\cite{spinodal} and 
suggests a coarsening in the stress field.
We indeed find a power-law growth
$R(t)\sim t^\phi$,
where $\phi=0.33\pm 0.01$ (cf. \cite{abj})
for all $s$ and $A$, as illustrated in Fig.~1.
This universal behavior and the value of $\phi$
agree with spinodal decomposition.

Now we present the evidence showing the system
in the long-time limit is stationary and critical
in the variables relevant to the dynamics.
Fig.~1 shows the first evidence of stationarity in
$R(t)$, where it reaches a plateau after transcient.
While the saturation rate depends on $s$ and $A$,
the value of the plateau $R_0$ depends only on $s$
due to $sL$ being the length the system has to contract to reach
a stress-free state. But this cannot be achieved due to PBC,
so that the blocks wind up in a frustrated state correlated
over this distance.
For fixed $L$, we verify that $R_0(s,L)$ increases linearly in $s$,
except very close to $s=0$ or 1.
The plateau extends up to ten decades in $t$,
until reaching the limit of numerical accuracy
(i.e., $10^{-15}$ in $F$, using double precision).
Further evidence
comes from $\bar F(t)/F_s$ which measures the effective
``distance'' from the instability limit for the system.
The stationary fluctuations of this ratio about a finite constant, as shown
in Fig.~2, is an important characteristic of a dynamical steady state
in SOC models\cite{btw,ofc,Others,Threshold,fault}.

Next, we show that the system is critical.
Firstly, for fixed $s$, we find
$R_0(s,L) \propto L$,
implying long-range correlations in the stress field.
More importantly,
$R$ satisfies finite-size scaling
$R(t,L)=t^\phi \tilde R(t/L^{1/\phi})$ (see Fig.~3),
with the asymptotic behavior
$\tilde R(x\to 0)={\rm const}$ and $\tilde R(x\to\infty)\propto x^{-\phi}$.
This is a clear signature of criticality.

Secondly, the avalanches are characterized by power-law
distributions, another hallmark of criticality.
From (\ref{force}), the force drop in one block-slip is given by
$F^{\rm slip}=K u/\alpha$\cite{ofc,gofc},
where $u$ denotes the slip distance.  Since $F^{\rm slip}\agt F_s$,
$u$ diminishes as $1/K$ when
the blocks gradually converge to a regular lattice.
An avalanche consists of a correlated sequence of $S$ block-slips.
Analogous to seismology\cite{scholz}, the ``seismic moment'' $M$ and
the ``radiated seismic energy'' $E$ can be computed:
\begin{eqnarray}
M&=& K \sum_{l=1}^S u_l = \alpha\sum_{l=1}^S F_l^{\rm slip}
\agt  \alpha F_s S,
\label{moment}\\
E&=& {1\over 2} \sum_{l=1}^S F^{\rm slip}_l u_l
={\alpha\over 2K} \sum_{l=1}^S (F_l^{\rm slip})^2
\agt {F_s M\over 2 K} .
\label{energy}
\end{eqnarray}
These can be measured experimentally, as is precisely done
for earthquakes.
In our context, $E$ equals the energy dissipated by friction.
We find that the distribution of $S$  approaches a power law
$P(S)\sim S^{-(B+1)}$ with $B=0.15\pm 0.05$ (see Fig.~4),
when $R(t,L)\to R_0(L)$ for $t \agt L^{1/\phi}$.
Consequently, both distributions of $M$ and the scaled energy $EK$
are stationary and follow the same power law.
From these results, we conclude that our system
approaches in a punctuated manner
without ever reaching its final equilibrium,
and is driven into a marginally stable state
which is critical in physical variables.

The approach to thermodynamic equilibrium via
spinodal decomposition\cite{spinodal} differs from our case
in that its dynamics is governed by thermal fluctuations,
not a threshold.
On the other hand,
it is well known that all
previous SOC models \cite{btw,ofc,Others,Threshold,fault} constructed
with threshold dynamics do not exhibit SOC with PBC, because
the periodic boundaries forbid an outflux of
the conserved quantity  (the total number of grains or the total
force on the array) to balance its external influx.
The key difference with our model is that
the variables $F_x$ and $F_y$ which are redistributed conservatively
can be of either sign,
with the total force remaining at zero during both block slips
and loading,
so that $F$ which determines topplings (i.e., slips)
is neither monotonic nor conserved.
This separation of the conservation law from the toppling condition
allows for criticality despite PBC.
Another way to rationalize our results is
to notice that, as mentioned above,
the mechanisms at work to produce powerlaws in the presence of
multiplicative noise
(amplification by multiplication followed by reinjection) \cite{Solomon}
might also be relevant here.

We also test the robustness of the critical properties against
boundary effects.
We find the same exponent $B$ within numerical uncertainty
for PBC, FBC, and CBC.
This is remarkable in view of their different
equilibrium states and paths of approach,
and the usual sensitive dependence of SOC systems
on boundary conditions.

Our model depends significantly on only one
factor: the conservation. To show this,
we introduce non-conservation
by adding a term $-(x,y)\kappa K $ to $(F_x,F_y)$ in (\ref{force}),
which may represent
harmonic couplings of each block with another surface,
as in earthquake models\cite{ofc}.
The parameter $\kappa$ then quantifies the level of non-conservation,
with $\alpha$ in (\ref{slip}) replaced by
$1/(2+2s+\kappa)$\cite{ofc,gofc}.
For not too small $\kappa (\agt 0.1),  \bar F(t)/F_s$ does not
show any stationary regime, and
$P(S)$ is exponential
for all choices of sampling intervals in $t$.
Consequently, the SOC state for $\kappa =0$ is lost except at most
for a small interval near $\kappa=0$.

\underline{\sl Conclusion:}
It has been suggested that SOC in sandpile and spring-block models arises
from a
de-sychronization mechanism that is initiated by inhomogeneities from a
free boundary \cite{Threshold}.
We have studied a system which shows SOC without such inhomogeneities.
This reveals a different mechanism whereby
correlations (or ``self-organized'' regions) 
gradually build up {\em in the bulk\/}
via a coarsening process.
The completion time $\tau \approx L^{1/\phi}$ (cf. Fig.~3)
is characterized by the coarsening exponent $\phi$,
which should be compared with the invasion time in previous models.
The associated power-law exponent is extremely robust
(universal)  with respect to
the initial disorder, the initial
strain, the size of the system and the type of boundary conditions.
While exhibiting a well-defined steady state in the
characteristic length and the average force, the model has also a transient
nature when viewed in the slip distance
toward the equilibrium
and in the sweeping of the spring coefficient $K$. This teaches us that
experimental
systems that appear transiently driven might in fact be stationary in the
variable relevant to the dynamics, especially when converging to a
fundamental
equilibrium state. Search schemes and optimization techniques using the
sweeping
of a control parameter such as in simulated annealing to get access to the
fundamental state or to the optimal solution might exhibit this kind of
phenomenon
in which the relaxation is characterized by a wide distribution of jumps.

The authors wish to thank Henrik Jeldtoft Jensen
for useful discussions.
J.V.A. wishes to acknowledge supports from
the European Union Human Capital and Mobility Program
contract number ERBCHBGCT920041 under the direction of Prof. E. Aifantis.
K.-t.L. is supported by the National Science Council and the NCHC of ROC.
D.S. is partially supported  by NSF EAR9615357.



\begin{figure}[htp]
\epsfig{figure=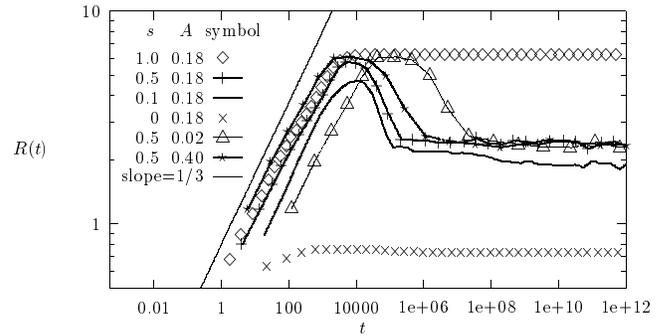,height=1.7in,angle=-0}
\caption{
Characteristic length $R(t)$ vs time $t\equiv K$
for different sets of $s$ and $A$, showing the universal 1/3 power law
and plateau. $L=40$ with PBC.
}
\label{fig1}
\end{figure}

\begin{figure}[htp]
\epsfig{figure=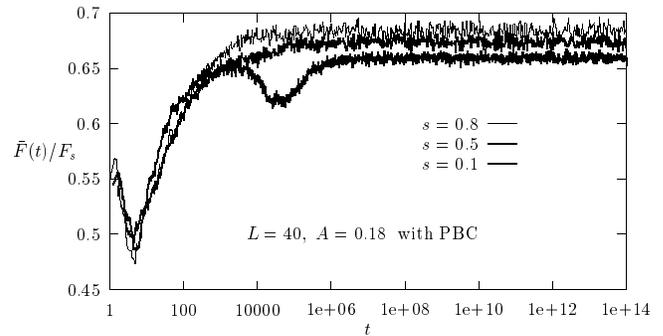,height=1.7in,angle=-0}
\caption{
Spatial averaged force on a block normalized by the threshold
vs time, showing approaches to stationarity.
}
\label{fig2}
\end{figure}

\begin{figure}[htp]
\epsfig{figure=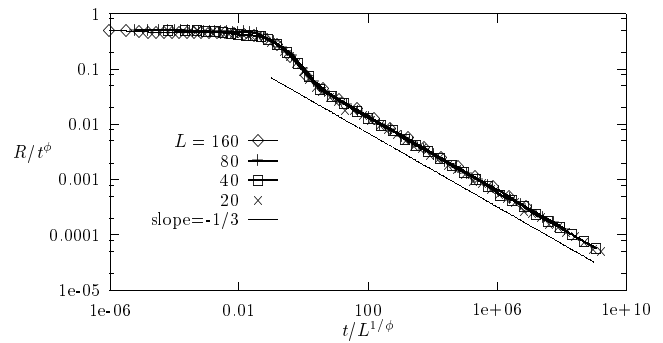,height=1.7in,angle=-0}
\caption{
Finite-size scaling plot of the characteristic length in the
stress field, $R(t,L)$.
$s=0.5$, $A=0.18$ with PBC.
}
\label{fig3}
\end{figure}

\begin{figure}[htp]
\epsfig{figure=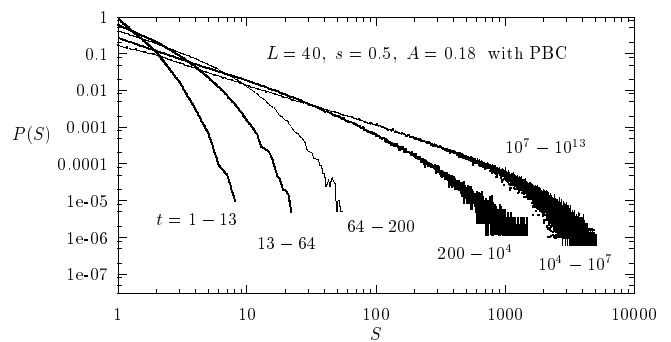,height=1.7in,angle=-0}
\caption{
$P(S)$ obtained from successive intervals in $t$
to show the approach to power law as $R(t)\to R_0$.
}
\label{fig4}
\end{figure}



\begin{references}

\bibitem{Strogatz} S.H. Strogatz and I. Steward, Scientific American {\bf
269},
102 (Dec, 1993).

\bibitem{Hopfield} A.V.M. Herz and J.J. Hopfield, Phys. Rev. Lett. {\bf
75}, 1222 (1995).

\bibitem{btw} P. Bak, C. Tang, K. Wiesenfeld, Phys. Rev. A {\bf 38}, 364
(1988);
Dhar D. et Ramaswamy R., Phys.Rev.Lett. {\bf 63}, 1659 (1989); Dhar
D., {\it ibid.} {\bf 64}, 1613 (1990).

\bibitem{ofc} Z. Olami, J. S.
Feder, and K. Christensen, Phys. Rev. Lett. {\bf 68}, 1244 (1992);
K. Christensen and Z. Olami, Phys. Rev. A {\bf 46}, 1829 (1992).

\bibitem{Others}
H. Feder and J. Feder, Phys. Rev. Lett., {\bf 66}, 2669 (1991);
I.M. J\'anosi and J. Kert\'esz, Physica A {\bf 200}, 179 (1993);
J. E. S. Socolar, G. Grinstein and C. Jayaprakash, Phys. Rev. E {\bf 47},
2366 (1993); P. Grassberger, Phys. Rev. E., {\bf 49}, 2436 (1994).

\bibitem{Threshold} K. Christensen, PhD Thesis
Oslo,  (Nov. 1992); A. Corral, C.J. P\'erez, A. D\'iaz-Guilera and A.
Arenas, Phys.Rev.Lett. {\bf 74}, 118 (1995); A.A. Middleton and C. Tang,
{\it ibid.} {\bf 74}, 742 (1995); S. Bottani, {\it ibid.}
{\bf 74}, 4189 (1995); L. Gil and D. Sornette, {\it ibid.}
{\bf 76}, 3991 (1996); A. Corral, C.J. P\'erez, A. D\'iaz-Guilera,
{\it ibid.} {\bf 78}, 1492 (1997).

\bibitem{fault} D. Sornette, P. Miltenberger and C.
Vanneste, Pageoph {\bf 142}, 491 (1994); in ``Recent Progresses in
Statistical
Mechanics and Quantum Field Theory'', eds. P. Bouwknegt et al., (World
Scientific,
Singapore, 1995), p.313-332.

\bibitem{Paczuski} M. Paczuski, S. Maslov and P. Bak, Phys. Rev. E {\bf
53}, 414 (1996).

\bibitem{Sor} D. Sornette and I. Dornic, Phys.Rev. E {\bf 54}, 3334 (1996);
D. Sornette, A. Johansen and I. Dornic, J.Phys.I France {\bf 5}, 325
(1995).

\bibitem{Cannelli} Cannelli G., Cantelli R. and Cordero F.,
Phys.Rev.Lett. {\bf 70}, 3923 (1993).

\bibitem{Clauss}
W. Clauss {\it et al}, Europhys. Lett. {\bf 12}, 423 (1990).

\bibitem{Barkhausen} P.J. Cote and L.V. Meisel,
 Phys. Rev. Lett. {\bf 67} 1334 (1991); J.S. Urbach, R.C. Madison and J.T.
Markert,
{\it ibid.} {\bf 75} 276 (1995).

\bibitem{Sweeping}  D. Sornette, Phys. Rev. Lett. {\bf 72} 2306 (1994);
J. Phys. I {\bf 4}, 209 (1994).

\bibitem{Sethna} O. Perkovic, K. Dahmen and J.P. Sethna, Phys. Rev. Lett.
{\bf 75} 4528 (1995).

\bibitem{Vilotte} J. Schmittbuhl, J.P. Vilotte and S. Roux, J. Geophys.
Res. {\bf 101}, 27741 (1996).

\bibitem{Solomon} R. Graham and A. Schenzle, Phys. Rev. A {\bf 25},
1731 (1982); A. Schenzle and H. Brand, Phys. Rev. A {\bf 20},
1628 (1979); Levy, M. and Solomon, Int. J. Mod. Phys.
C {\bf 7}, 65 (1996); Sornette, D. and Cont, R., J. Phys. I France
{\bf 7}, 431 (1997).

\bibitem{gofc} K.-t. Leung, J. M\"uller, and J. V. Andersen, J Phys I
France {\bf 7},  423 (1997).

\bibitem{footnote1}
Although our model is driven by increasing $K$ and fixed $F_s$,
physically only an increasing $K/F_s$ matters.
Thus it describes
the general behavior of real materials which
become weaker (smaller $F_s$) as they get stiffer (larger $K$).

\bibitem{expt}
A. Groisman and E. Kaplan, Europhys. Lett. {\bf 25}, 415 (1994).

\bibitem{abj}
J. V. Andersen, Y. Brechet, and H. J. Jensen, Europhys. Lett. {\bf 26}, 13
(1994);
J. V. Andersen, Phys. Rev. B {\bf 49}, 9981 (1994);
K.-t. Leung and J. V. Andersen, Europhys. Lett. (in press, 1997).

\bibitem{spinodal}
See, e.g., J.D. Gunton, M. San Miguel and P.S. Sahni,
{\it Phase Transitions and Critical Phenomena\/}, Vol. 8,
ed. C. Domb and J.L. Lebowitz (Academic, NY, 1983).

\bibitem{scholz}
See, e.g., C.H. Scholz, {\it The Mechanics of Earthquakes and Faulting\/}
(Cambridge, Cambridge, 1990).



\end{references}
\end{document}